\documentclass[showpacs, amsmath, amssymb, 
 pra,twocolumn]{revtex4-1} 
\usepackage{amsmath}
\usepackage{amssymb}

\newcommand{\braket}[2]{\left\langle #1 | #2\right\rangle}
\newcommand{\ketbra}[2]{\left| #1 \right \rangle\! \left \langle #2\right|}
\newcommand{\bra}[1]{\left\langle #1 \right|}
\newcommand{\ket}[1]{\left| #1 \right\rangle}

\begin{document}

\title{Majorization relations for a set of two-mode squeezed number states}
\author{Ryo Namiki}
\affiliation{Department of Physics, Gakushuin University, 1-5-1 Mejiro, Toshima-ku,  Tokyo 171-8588, Japan}

\date{\today}

\begin{abstract}
Two-mode squeezed number states (TMSNS)  are natural generalization of two-mode squeezed vacuum states. It has been known that every TMSNS is entangled whenever the squeezing parameter is non-zero. For a pair of entangled pure states Nielsen's majorization theorem tells us whether one state can be transformed into the other state through local operation and classical communication based on the majorization property on their probability distributions of Schmidt bases.  
In this report we find two examples of majorization relations for a set of TMSNS.
\end{abstract}
 

\maketitle

\noindent {\bf 1. Introduction}\\
Given a pair of physical states it is fundamental to determine whether two states can be converted each other or there is some (dis)order in conversion associated with a certain set of physical quantities. Such a question plays an essential role  in thermal physics and information sciences. 
In quantum information science, the convertibility between maximally entangled states by local unitary operations is intensively used as a basic tool to accomplish quantum communication and computation tasks \cite{NC00}. Moreover, the  convertibility under local operation and classical communications (LOCC)  characterizes  the strength of quantum entanglement  \cite{Horo09}.

Nielsen's majorization theorem \cite{Nielsen99} states that an entangled pure state can be transformed into another entangled pure state by an LOCC protocol 
iff the probability distribution of Schmidt basis is majorized by that of the target state. This theorem was extended for pure states in infinite dimensional systems \cite{Owari04} (See also \cite{Asakura16}). 

In an infinite dimensional bipartite system, a two-mode squeezed state is often thought to play the role of a maximally entagle state. A wider class of entagled pure states that contains the two-mode squeezed state is the set of two-mode number states (TMSNS). The TMSNS are natural generalization of two-mode squeezed states  \cite{CHIZHOV199333}. They are entangled pure states and form an orthonormal basis  \cite{Namiki10J}.  
Interestingly, an infinite sequence of ordered convertibility can be found between a class of TMSNS \cite{Garc12}. 
The sequence concerns the set of TMSNSs whose probability distributions in Schmidt basis take the form of the negative binomial distribution \cite{AgarwalNBS,Namiki10J}. However, it is a rather limited case that a TMSNS has such a property in the probability distribution. Hence, it is natural to ask whether there exists such a LOCC convertibility for a wider class of TMSNSs. In this report, we find two examples of the majorization relations for a set of TMSNSs.

\noindent {\bf 2. Majorization and LOCC conversion between bipartite states}\\
Let us consider the form of $d$-dimension probability distribution $\mathbf{p} =(p_1,p_2,..., p_d)^T $ 
satisfying $ \sum_{i}p_i=1$ and $p_i \ge 0$. 
For such forms, we say $\mathbf{p}$ majorizes $\mathbf{q} $ if it holds
\begin{align}
 \sum_{n=1}^m{p_i^\downarrow} \ge  \sum_{n=1}^m{q_i^\downarrow}    \quad \forall m <d, \label{MojoM1}
\end{align} where the symbol $^\downarrow $ means that the elements are rearranged in the decreasing order $p_1^\downarrow  \ge p_2^\downarrow \ge \cdots \ge p_d^\downarrow $. We denote the majorization condition as
\begin{align}
\mathbf{p} \succ \mathbf{q}  \quad \Longleftrightarrow \quad  \sum_{n=1}^m{p_i^\downarrow} \ge  \sum_{n=1}^m{q_i^\downarrow}  \quad \forall m <d.
\end{align}
We say that a pure entangled state $\ket{\phi_{\mathbf{p}}}$ majorizes a pure entangled state $\ket{\phi_{\mathbf{q}}}$ if there exists local unitary operators $u, u^\prime , v, v^\prime $ and orthonormal bases $\{\ket{e_i} \},  \{\ket{f_i} \} $ in such a way that they have Schmidt forms 
\begin{align}
 u_A \otimes  u_B^\prime \ket{\phi_{\mathbf{p}} } =&  \sum_{i} \sqrt{p_i} \ket{e_i}\otimes  \ket{f_i}   ,  \nonumber \\  v_A \otimes  v_B^\prime \ket{\phi_{\mathbf{q}} }  =&  \sum_{i} \sqrt{q_i} \ket{e_i}\otimes  \ket{f_i}, 
\end{align} and the majorization relation Eq.~\eqref{MojoM1} holds for the coefficients $\{p_i\}$ and $\{q_i\}$. 
We may also denote this majorization condition for quantum states as
\begin{align}
\ket{\phi_{\mathbf{p}} } \succ \ket{\phi_{\mathbf{q}} }  \quad \Longleftrightarrow \quad  \sum_{n=1}^m{p_i^\downarrow}  \ge  \sum_{n=1}^m{q_i^\downarrow} \quad \forall m <d.
\end{align}
Neilsen's theorem proves that an LOCC process can convert $\ket{\phi_{\mathbf{q}} } $ into $\ket{\phi_{\mathbf{p}} }$ if and only if  $\ket{\phi_{\mathbf{p}} } \succ \ket{\phi_{\mathbf{q}} } $ holds for finite dimensional states $d < \infty$. We may write symbolically, 
\begin{align}
\ket{\phi_{\mathbf{q}} } \overset{\text{LOCC}}{\longrightarrow}  \ket{\phi_{\mathbf{p}} }  \quad \Longleftrightarrow \quad  \ket{\phi_{\mathbf{p}} } \succ \ket{\phi_{\mathbf{q}} }.
\end{align}

\noindent {\bf 3. A class of infinite dimensional pure entangled states}\\

Let us write number states $\ket{n}$ and assume ordinary commutation relations for annihilation and creation operators of two modes $[a,a^\dagger] =[b, b^\dagger]=i$ and $[a, b]=[a, b^\dagger]=0$. Let be $\lambda \in (0,1)$ and  define coupled annihilation operators as 
\begin{eqnarray}
 \hat A_\lambda :=  \frac{a  -\lambda b ^\dagger }{\sqrt{1-\lambda^2}}, \quad  \label{defA}
\hat B_\lambda:=\frac{b  -\lambda a^\dagger}{\sqrt{1-\lambda ^2}}. \end{eqnarray}It implies the commutation relations 
 $[\hat A, \hat A^\dagger] =[\hat B, \hat B^\dagger]=i$ and $[\hat A,\hat  B]=[\hat A, \hat B^\dagger]=0$. 
Let us define the vacuum $\ket{\psi_{0,0}} $ assocated with $\hat A_\lambda$ and $ \hat  B_\lambda$ as the state which satisfies 
\begin{align}
\hat  A_\lambda \ket{\psi_{0,0}(\lambda ) } =0, \quad  \hat  B_\lambda \ket{\psi_{0,0}(\lambda ) } =0. 
\end{align}
This relation gives the familiar form of the two mode squeezed vacuum 
\begin{align}
 \ket{\psi_{0,0}(\lambda ) }=  \sqrt{1-\lambda^2} \sum_{n=0}^\infty  \lambda ^ n   \ket{n,n}. 
\end{align}
 TMSNSs can be defined as the simultaneous eigenstates associated with the coupled number operators $\hat  N_A = A_\lambda ^\dagger \hat  A_\lambda $ and  $\hat  N_B = \hat  B_\lambda ^\dagger \hat B_\lambda $:
\begin{align}
|\psi_{N_A,N_B} (\lambda) \rangle =\frac{ (\hat A_\lambda^\dagger)^{N_A} }{\sqrt{N_A!}} \frac{(\hat B_\lambda^\dagger)^{N_B}}{\sqrt{N_B!}}  \ket{\psi_{0,0}(\lambda ) }. \label{basisi}
\end{align}
We can find the Schmidt decomposed form of the TMSNSs \cite{Namiki10J}
\begin{align}
& |\psi_{N_A,N_B} (\lambda) \rangle  \nonumber \\ 
 =&  \sum_{m=0}^{\infty }  C_m(N_A,N_B, \lambda )  |N_A-N_B+m \rangle_A |m \rangle_B ,\end{align}
where we assume $N_A \ge N_B$ and the Schmidt coefficients $\{ C_m\}_{m=0,1,2,...}$  are given by   
\begin{align}
 & 
C_m(N_A,N_B, \lambda ) \nonumber\\
  =& (1-\lambda^2)^{\frac{N_A-N_B+1}{2}}  \sum_{k=0}^{\min\{m,N_B\}}\Bigg( (1-\lambda^2)^k ( -\lambda )^{N_B-k} \nonumber  \\
   &\times   \lambda^{m-k} \frac{\sqrt{N_A! N_B! (N_A-N_B+m)! m!} }{k!(m-k)!(N_A-N_B +k)!(N_B-k)!} \Bigg) .\label{QnSchmidtC} 
  \end{align}
  
For the class of TMSNSs involving single-mode excitations, e.g., $N_B=0$, 
 it was shown \cite{Garc12} that 
\begin{align}
|\psi_{n,0} (\lambda) \rangle  \succ |\psi_{n+m,0} (\lambda) \rangle , 
\label{QnM00}
\end{align} holds for $n,m=0,1,2,3,...$. Hence, a complete set of ordered  convertible property under LOCC was established in this class of TMSNSs
\begin{align}
|\psi_{n+m,0} (\lambda) \rangle   \overset{\text{LOCC}}{\longrightarrow} |\psi_{n,0} (\lambda) \rangle.  
\label{QnM01}
\end{align}
To reach the majorization relation Eq.~\eqref{QnM00} the following Lemma is essential:

\textbf{Lemma. (Lemma 3.1 of \cite{Markus64})}---  An existence of a column-stochastic matrix $D$ that satisfies 
\begin{align}
 \mathbf{q} = D\mathbf{p}, 
\end{align} is sufficient to the majorization condition
\begin{align}
 \mathbf{p} \succ \mathbf{q}, 
\end{align} for positive infinite sequences  $\mathbf{p}, \mathbf{q}$. 
Here, a positive real matrix $D$ ($D_{i.j} \ge 0$ for all $ i,j $) is called column-stochastic if its 
colum sum is one ($\sum_{i}D_{i,j} =1$ for all $j$) and row-sum is less than one  ($\sum_{j}D_{i,j}  \le 1$ for all $i$). These relations suggest that $\mathbf{q}$ has a higher entropy than $\mathbf{p}$, and the two probability distributions, $\mathbf{p}$ and  $\mathbf{q}$,  are actually connected with 
  a stochastic transformation $D$. 

For a notation convention, let us write the probability distribution associated with the Schmidt basis of $ |\psi_{N_A,N_B} (\lambda) \rangle $ as 
\begin{align}
(\mathbf{p}_{N_A, N_B})_m:=& |\braket{N_A+N_B-m,m}{\psi_{N_A,N_B} (\lambda)}|^2
\nonumber \\ 
=& |C_m(N_A,N_B, \lambda)|^2     
\end{align} where the Schmidt coefficient $C_m$ is given in Eq.~\eqref{QnSchmidtC}. It was shown in  \cite{Garc12} 
\begin{align}
 \mathbf{p}_{n+m,0}=D^m \mathbf{p}_{n,0} \label{QnM02}
\end{align} is fulfilled with the following column-stochastic matrix
\begin{align}
D:=&(1-\lambda^2) \left(
\begin{array}{ccccc}
 1  & 0 & 0  &   \\
 \lambda^2   & 1 & 0 &    \ddots  \\
 \lambda^4& \lambda^2   & 1 &\ddots \\ 
    & \ddots & \ddots  \quad \ddots   &\ddots \\
 \end{array}
\right)\nonumber \\
 = & \left(
\begin{array}{ccccc}
 a_0  & 0 & 0  &  \cdots \\
 a_1 &  a_0 & 0 &    \cdots  \\
 a_2 &  a_1 & a_0   & \ddots   \\ 
    & \ddots & \ddots  \quad \ddots   &\ddots \\
 \end{array}
\right). \label{QnMatD}
\end{align}
for any $n,m =0,1,2,...$ and  $\lambda \in [0,1)$. The final expression is a form of the Toeplitz matrix. 
The relation Eq.~\eqref{QnM02} implies the majorization relation Eq.~\eqref{QnM00} due to the above Lemma. Then, an extension of Nielsen's theorem for infinite dimension implies the capability of the LOCC transformation Eq.~\eqref{QnM01}. 


\noindent {\bf 4. Main results}\\
In what follows, we find the majorization relations between the three TMSNSs $\psi_{0,0}$, $\psi_{1,0}$, and $\psi_{1,1}$. Our method has basically two steps: (i) We empirically find a Toeplitz matrix in the form of Eq.~\eqref{QnMatD} that connects the probability distributions of Schmidt bases. (ii) Then we determine a parametric regime of $\lambda $ where the matrix becomes column-stochastic. Thereby, the majorization condition is established in the specified regime. 

The squared Schmidt coefficients of the three states  $\psi_{1,1}$, $\psi_{1,0}$, and $\psi_{0,0}$ are respectively given by \begin{widetext} 
\begin{align}
\mathbf{p}_{11}=  (1-\lambda^2 )  \left(
\begin{array}{c}
 \lambda ^2 \\
 \left(1-2 \lambda ^2\right)^2 \\
 \lambda ^2 \left(2-3 \lambda ^2\right)^2 \\
 \lambda ^4 \left(3-4 \lambda ^2\right)^2 \\
 \vdots \\
 \lambda ^{2(n-1)}(n-(n+1)\lambda ^2 )^2 \\
 \vdots \\
\end{array}
\right),   \quad 
\mathbf{p}_{10}= (1-\lambda^2 )^2 \left(
\begin{array}{c}
 1 \\
 2\lambda ^2 \\
 3\lambda ^4 \\
 4\lambda ^6 \\
 \vdots \\
 (n+1)\lambda ^{2n} \\
 \vdots \\
 \end{array}
\right),  \quad 
\mathbf{p}_{00}= (1-\lambda^2 ) \left(
\begin{array}{c}
 1 \\
 \lambda ^2 \\
 \lambda ^4 \\
 \lambda ^6 \\
 \vdots \\
 \lambda ^{2n} \\
 \vdots \\
 \end{array}
\right)
\end{align}

Let us consider  $\psi_{1,1}$ and $\psi_{1,0}$. One may empirically find that 
their probability distributions fulfill
\begin{align}
\mathbf{p}_{11}=A \mathbf{p}_{10}
\end{align}
where $A$ is a Toeplitz matrix defined as follows:
\begin{align}
A=&
(1-\lambda^2)^{-1}\left(
\begin{array}{ccccc}
 \lambda ^2 \quad \quad & 0 & 0 & 0 & 0 \\
 2 \lambda ^4-4 \lambda ^2+1 & \lambda ^2 \quad \quad & 0 & 0 & 0 \\
 2 \lambda ^6-4 \lambda ^4+2 \lambda ^2 & 2 \lambda ^4-4 \lambda ^2+1 & \lambda ^2  \quad \quad & 0 & 0 \\
 2 \lambda ^8-4 \lambda ^6+2 \lambda ^4 & 2 \lambda ^6-4 \lambda ^4+2 \lambda ^2 & 2 \lambda ^4-4 \lambda ^2+1 & \lambda ^2 & 0 \\
\vdots  & \vdots & \vdots \quad \quad \ddots & &\ddots \\
 \end{array}
\right)  \nonumber \\ 
=&(1-\lambda^2)^{-1}\left(
\begin{array}{ccccc}
 \lambda ^2 & 0 & 0 & 0 & 0 \\
  2(1- \lambda ^2)^2 -1 & \lambda ^2 & 0 & 0 & 0 \\
 2 \lambda ^2(1- \lambda ^2)^2 & 2(1- \lambda ^2)^2 -1 & \lambda ^2   & 0 & 0 \\
2 \lambda ^4(1- \lambda ^2)^2  & 2 \lambda ^2(1- \lambda ^2)^2 &2(1- \lambda ^2)^2 -1& \lambda ^2 & 0 \\
\vdots  & \vdots & \vdots \quad \quad \ddots & &\ddots \\
 \end{array}
\right) 
. \label{QnMatA12}
\end{align}
\end{widetext}
From the expression of the first line we can readily see that the column sum is one ($\sum_i A_{i.j}=1$). From the second expression we can see that all elements are nonnegative if $2(1- \lambda ^2)^2 -1 \ge 0 $.  
This condition implies 
\begin{align}
 | \lambda |  \le  \lambda_{10 \succ 11}^{(0)} := \sqrt{\frac{2-\sqrt{2}}{2} } = 0.541196... \label{QnMain1C}
\end{align} 
If all elements are non negative, the row sum $s_i=\sum_j A_{i,j}$ is an increasing sequence and bounded by one (e.g., $s_i \le \sum_{i} A_{i,1}=1$). Therefore, $A $ is proven to be column-stochastic if $\lambda \in [0,  \lambda_{10 \succ 11}^{(0)} ] $. Hence,  from the Lemma, we can conclude that the following majorization relation holds 
\begin{align}
\left. \begin{array}{c}
 \psi_{10} (\lambda)   \\
 \psi_{01} (\lambda)  \\ \end{array} 
 \right\}
  \succ \psi_{11} (\lambda ) ,  \quad \lambda \in [0, \lambda_{10 \succ 11}^{(0)}]. \label{mainR1}
\end{align}

Let us consider  $\psi_{1,1}$ and $\psi_{0,0}$. We can see that the following Toeplitz matrix converts the probability distributions of  $\psi_{1,1}$ and $\psi_{0,0}$ as 
\begin{align}
\mathbf{p}_{11}=A \mathbf{p}_{00} \label{QnMain2}
\end{align}
where 
\begin{widetext}
\begin{align}
A=\left(
\begin{array}{ccccc}
 \lambda ^2 & 0 & 0 & 0 & \cdots \\
 3 \lambda ^4-4 \lambda ^2+1 & \lambda ^2 & 0 & 0 &  \cdots  \\
 5 \lambda ^6-8 \lambda ^4+3 \lambda ^2 & 3 \lambda ^4-4 \lambda ^2+1 & \lambda ^2 & 0 &    \\
 7 \lambda ^8-12 \lambda ^6+5 \lambda ^4 & 5 \lambda ^6-8 \lambda ^4+3 \lambda ^2 & 3 \lambda ^4-4 \lambda ^2+1 & \lambda ^2 & \ddots \\
\vdots  & \vdots & \vdots \quad \quad \ddots & &\ddots \\
 \end{array}
\right) = \left(
\begin{array}{ccccc}
 a_0  & 0 & 0  &  \cdots \\
 a_1 &  a_0 & 0 &    \cdots  \\
 a_2 &  a_1 & a_0   & \ddots   \\ 
    & \ddots & \ddots  \quad \ddots   &\ddots \\
 \end{array}
\right). \label{QnMatA}
\end{align}
\end{widetext}
We can readily confirm that 
the  column-sum condition is satisfied ($ \sum_{i=1}^\infty  A_{i,j}=1 $). 
The matrix elements can be written as 
\begin{align}
A_{{i,i}}=   a_0(\lambda ) :=&\lambda ^2,  \nonumber \\
   A_{ n+i,i}=  a_n (\lambda ) :=& \lambda ^{2 (n-1)} (\lambda^4 +2n(1-\lambda^2)^2 -1 )
   \nonumber  \\ & \quad (n=1,2,3,...)
\end{align}
and 
 the following recurrence formula holds for $n \ge 1 $
\begin{align}
 a_{n+1}(\lambda )= \lambda^2  a_{n }(\lambda ) + 2 \lambda ^{2n} (1- \lambda )^2 . 
\end{align} This implies $a_{n+1}$ is positive if $a_n$ is positive. Hence, we can show all elements are positive if $a_1$ is positive. The condition $a_1(\lambda ) \ge 0$, namely, $3 \lambda ^4 -4 \lambda ^2 +1  \ge 0$
is fulfilled if 
\begin{align}
 | \lambda |  \le   \lambda_{00 \succ 11}^{(0)} := 1/\sqrt 3 = 0.57735...
\end{align} If all elements are non negative, the row sum $s_i=\sum_j A_{i,j}$ is an increasing sequence and bounded by one (e.g., $s_i \le \sum_{i} A_{i,1}=1$). 
Therefore, $A$ is column-stochastic if $\lambda \in [0, \lambda_{00 \succ 11}^{(0)} ] $. This implies 
\begin{align}
 \psi_{00} (\lambda)  \succ \psi_{11} (\lambda ), \quad \lambda \in [0, \lambda_{00 \succ 11}^{(0)}]. 
\end{align}

We can show that the majorization range on $\lambda$ 
becomes a bit wider by a modification of the matrix $A$. It holds $\mathbf{p}_{11}=A^\prime \mathbf{p}_{00} $ with  
\begin{widetext}
\begin{align}
A^\prime=  \left(
\begin{array}{cccccc}
 \lambda ^2-\lambda ^4 & \lambda ^2-\lambda ^4 & \lambda ^2 & 0 & 0 & \cdots \\
 4 \lambda ^4-4 \lambda ^2+1 & 0 & 0 & 0 & 0 & 0 \\
 5 \lambda ^6-8 \lambda ^4+3 \lambda ^2 & 4 \lambda ^4-4 \lambda ^2+1 & 0 & 0 &  \ddots &  \vdots\\
 7 \lambda ^8-12 \lambda ^6+5 \lambda ^4 & 5 \lambda ^6-8 \lambda ^4+3 \lambda ^2 & 4 \lambda ^4-4 \lambda ^2+1 & 0 & 0 &  \\
 9 \lambda ^{10}-16 \lambda ^8+7 \lambda ^6 & 7 \lambda ^8-12 \lambda ^6+5 \lambda ^4 & 5 \lambda ^6-\underline{9 \lambda ^4}+3 \lambda ^2 & 4 \lambda ^4-\underline{3 \lambda ^2}+1 & 0 & 0 \\
 11 \lambda ^{12}-20 \lambda ^{10}+9 \lambda ^8 & 9 \lambda ^{10}-16 \lambda ^8+7 \lambda ^6 & 7 \lambda ^8-12 \lambda ^6+5 \lambda ^4 & 5 \lambda ^6-\underline{9 \lambda ^4}+3 \lambda ^2 & 4 \lambda ^4- \underline{3 \lambda ^2} +1 & 0 \\
 \vdots  & \ddots  & \ddots   & \ddots  & \ddots  & \ddots  \\
\end{array}
\right). 
\end{align}
\end{widetext}
Here, the underlines remark the modified elements possibly to be oversight. 
We can routinely show that $A^\prime$ is column-stochastic 
when 
\begin{align}
 | \lambda| \le \lambda_{00 \succ 11}^{(1)}:= \sqrt\frac{9-\sqrt {21}}{10} =0.6646... \label{QnMain2C}
\end{align}  This condition comes from $ {A^\prime }_{5,3}= 5 \lambda ^4- {9 \lambda ^2}+3 \ge 0$. 
Therefore, the majorization relation still holds for this regime: 
\begin{align}
 \psi_{00} (\lambda)  \succ \psi_{11} (\lambda ), \quad \lambda \in [0, \lambda_{00 \succ 11}^{(1)}]. \label{mainR2}
\end{align}

\noindent {\bf 5. Conclusion and Remarks}\\
We have found two examples of majorization relations for the TMSNSs,  $\psi_{1,1}$, $\psi_{1,0}$, and $\psi_{0,0}$   
  (Eqs.~\eqref{mainR1} and \eqref{mainR2} with the constraints on the squeezing parameter Eq.~\eqref{QnMain1C} and \eqref{QnMain2C}, respectively). 
Our approach was mostly heuristic, and seems scarcely be helpful to reach a general theorem.
It is desirable to find a systematic method to determine the majorization condition over a general pair of TMSNSs.

This work was supported by the ImPACT Program of Council for Science, Technology and Innovation (Cabinet Office, Government of Japan),  JSPS KAKENHI (Grant No. JP18H01157 and Grant No. JP18H05237), and
Cross-ministerial Strategic Innovation Promotion Program (SIP) (Council for Science, Technology and Innovation (CSTI)).


%


\begin{thebibliography}{10}%
\makeatletter
\providecommand \@ifxundefined [1]{%
 \@ifx{#1\undefined}
}%
\providecommand \@ifnum [1]{%
 \ifnum #1\expandafter \@firstoftwo
 \else \expandafter \@secondoftwo
 \fi
}%
\providecommand \@ifx [1]{%
 \ifx #1\expandafter \@firstoftwo
 \else \expandafter \@secondoftwo
 \fi
}%
\providecommand \natexlab [1]{#1}%
\providecommand \enquote  [1]{``#1''}%
\providecommand \bibnamefont  [1]{#1}%
\providecommand \bibfnamefont [1]{#1}%
\providecommand \citenamefont [1]{#1}%
\providecommand \href@noop [0]{\@secondoftwo}%
\providecommand \href [0]{\begingroup \@sanitize@url \@href}%
\providecommand \@href[1]{\@@startlink{#1}\@@href}%
\providecommand \@@href[1]{\endgroup#1\@@endlink}%
\providecommand \@sanitize@url [0]{\catcode `\\12\catcode `\$12\catcode
  `\&12\catcode `\#12\catcode `\^12\catcode `\_12\catcode `\%12\relax}%
\providecommand \@@startlink[1]{}%
\providecommand \@@endlink[0]{}%
\providecommand \url  [0]{\begingroup\@sanitize@url \@url }%
\providecommand \@url [1]{\endgroup\@href {#1}{\urlprefix }}%
\providecommand \urlprefix  [0]{URL }%
\providecommand \Eprint [0]{\href }%
\providecommand \doibase [0]{http://dx.doi.org/}%
\providecommand \selectlanguage [0]{\@gobble}%
\providecommand \bibinfo  [0]{\@secondoftwo}%
\providecommand \bibfield  [0]{\@secondoftwo}%
\providecommand \translation [1]{[#1]}%
\providecommand \BibitemOpen [0]{}%
\providecommand \bibitemStop [0]{}%
\providecommand \bibitemNoStop [0]{.\EOS\space}%
\providecommand \EOS [0]{\spacefactor3000\relax}%
\providecommand \BibitemShut  [1]{\csname bibitem#1\endcsname}%
\let\auto@bib@innerbib\@empty
\bibitem [{\citenamefont {Nielsen}\ and\ \citenamefont {Chuang}(2000)}]{NC00}%
  \BibitemOpen
  \bibfield  {author} {\bibinfo {author} {\bibfnamefont {M.~A.}\ \bibnamefont
  {Nielsen}}\ and\ \bibinfo {author} {\bibfnamefont {I.~L.}\ \bibnamefont
  {Chuang}},\ }\href@noop {} {\emph {\bibinfo {title} {Quantum Computation and
  Quantum Information}}}\ (\bibinfo  {publisher} {Cambridge University Press},\
  \bibinfo {year} {2000})\BibitemShut {NoStop}%
\bibitem [{\citenamefont {Horodecki}\ \emph {et~al.}(2009)\citenamefont
  {Horodecki}, \citenamefont {Horodecki}, \citenamefont {Horodecki},\ and\
  \citenamefont {Horodecki}}]{Horo09}%
  \BibitemOpen
  \bibfield  {author} {\bibinfo {author} {\bibfnamefont {R.}~\bibnamefont
  {Horodecki}}, \bibinfo {author} {\bibfnamefont {P.}~\bibnamefont
  {Horodecki}}, \bibinfo {author} {\bibfnamefont {M.}~\bibnamefont
  {Horodecki}}, \ and\ \bibinfo {author} {\bibfnamefont {K.}~\bibnamefont
  {Horodecki}},\ }\href {\doibase 10.1103/RevModPhys.81.865} {\bibfield
  {journal} {\bibinfo  {journal} {Rev. Mod. Phys.}\ }\textbf {\bibinfo {volume}
  {81}},\ \bibinfo {pages} {865} (\bibinfo {year} {2009})}\BibitemShut
  {NoStop}%
\bibitem [{\citenamefont {Nielsen}(1999)}]{Nielsen99}%
  \BibitemOpen
  \bibfield  {author} {\bibinfo {author} {\bibfnamefont {M.~A.}\ \bibnamefont
  {Nielsen}},\ }\href {\doibase 10.1103/PhysRevLett.83.436} {\bibfield
  {journal} {\bibinfo  {journal} {Phys. Rev. Lett.}\ }\textbf {\bibinfo
  {volume} {83}},\ \bibinfo {pages} {436} (\bibinfo {year} {1999})}\BibitemShut
  {NoStop}%
\bibitem [{\citenamefont {Owari}\ \emph {et~al.}(2008)\citenamefont {Owari},
  \citenamefont {Braunstein}, \citenamefont {Nemoto},\ and\ \citenamefont
  {Murao}}]{Owari04}%
  \BibitemOpen
  \bibfield  {author} {\bibinfo {author} {\bibfnamefont {M.}~\bibnamefont
  {Owari}}, \bibinfo {author} {\bibfnamefont {S.~L.}\ \bibnamefont
  {Braunstein}}, \bibinfo {author} {\bibfnamefont {K.}~\bibnamefont {Nemoto}},
  \ and\ \bibinfo {author} {\bibfnamefont {M.}~\bibnamefont {Murao}},\
  }\href@noop {} {\bibfield  {journal} {\bibinfo  {journal} {Quantum
  Information \& Computation}\ }\textbf {\bibinfo {volume} {8}},\ \bibinfo
  {pages} {30} (\bibinfo {year} {2008})}\BibitemShut {NoStop}%
\bibitem [{\citenamefont {Asakura}(2016)}]{Asakura16}%
  \BibitemOpen
  \bibfield  {author} {\bibinfo {author} {\bibfnamefont {D.}~\bibnamefont
  {Asakura}},\ }\href@noop {} {\bibfield  {journal} {\bibinfo  {journal} {RIMS
  Kokyuroku No.2033}\ } (\bibinfo {year} {2016})}\BibitemShut {NoStop}%
\bibitem [{\citenamefont {Chizhov}\ and\ \citenamefont
  {Murzakhmetov}(1993)}]{CHIZHOV199333}%
  \BibitemOpen
  \bibfield  {author} {\bibinfo {author} {\bibfnamefont {A.}~\bibnamefont
  {Chizhov}}\ and\ \bibinfo {author} {\bibfnamefont {B.}~\bibnamefont
  {Murzakhmetov}},\ }\href {\doibase
  https://doi.org/10.1016/0375-9601(93)90312-N} {\bibfield  {journal} {\bibinfo
   {journal} {Physics Letters A}\ }\textbf {\bibinfo {volume} {176}},\ \bibinfo
  {pages} {33 } (\bibinfo {year} {1993})}\BibitemShut {NoStop}%
\bibitem [{\citenamefont {Namiki}(2010)}]{Namiki10J}%
  \BibitemOpen
  \bibfield  {author} {\bibinfo {author} {\bibfnamefont {R.}~\bibnamefont
  {Namiki}},\ }\href {\doibase 10.1143/JPSJ.79.013001} {\bibfield  {journal}
  {\bibinfo  {journal} {J. Phys. Soc. Jpn.}\ }\textbf {\bibinfo {volume}
  {79}},\ \bibinfo {pages} {013001} (\bibinfo {year} {2010})}\BibitemShut
  {NoStop}%
\bibitem [{\citenamefont {Garc\'{\i}a-Patr\'on}\ \emph
  {et~al.}(2012)\citenamefont {Garc\'{\i}a-Patr\'on}, \citenamefont
  {Navarrete-Benlloch}, \citenamefont {Lloyd}, \citenamefont {Shapiro},\ and\
  \citenamefont {Cerf}}]{Garc12}%
  \BibitemOpen
  \bibfield  {author} {\bibinfo {author} {\bibfnamefont {R.}~\bibnamefont
  {Garc\'{\i}a-Patr\'on}}, \bibinfo {author} {\bibfnamefont {C.}~\bibnamefont
  {Navarrete-Benlloch}}, \bibinfo {author} {\bibfnamefont {S.}~\bibnamefont
  {Lloyd}}, \bibinfo {author} {\bibfnamefont {J.~H.}\ \bibnamefont {Shapiro}},
  \ and\ \bibinfo {author} {\bibfnamefont {N.~J.}\ \bibnamefont {Cerf}},\
  }\href {\doibase 10.1103/PhysRevLett.108.110505} {\bibfield  {journal}
  {\bibinfo  {journal} {Phys. Rev. Lett.}\ }\textbf {\bibinfo {volume} {108}},\
  \bibinfo {pages} {110505} (\bibinfo {year} {2012})}\BibitemShut {NoStop}%
\bibitem [{\citenamefont {Agarwal}(1992)}]{AgarwalNBS}%
  \BibitemOpen
  \bibfield  {author} {\bibinfo {author} {\bibfnamefont {G.~S.}\ \bibnamefont
  {Agarwal}},\ }\href {\doibase 10.1103/PhysRevA.45.1787} {\bibfield  {journal}
  {\bibinfo  {journal} {Phys. Rev. A}\ }\textbf {\bibinfo {volume} {45}},\
  \bibinfo {pages} {1787} (\bibinfo {year} {1992})}\BibitemShut {NoStop}%
\bibitem [{\citenamefont {Markus}(1964)}]{Markus64}%
  \BibitemOpen
  \bibfield  {author} {\bibinfo {author} {\bibfnamefont {A.~S.}\ \bibnamefont
  {Markus}},\ }\href {http://stacks.iop.org/0036-0279/19/i=4/a=R02} {\bibfield
  {journal} {\bibinfo  {journal} {Russian Mathematical Surveys}\ }\textbf
  {\bibinfo {volume} {19}},\ \bibinfo {pages} {91} (\bibinfo {year}
  {1964})}\BibitemShut {NoStop}%
\end{thebibliography}
\end{document}